\documentstyle[art10,hip-100ps,twoside,epsf]{article}
\begin{document}
\cimo{233}{240} \setcounter{page}{233} 
\thispagestyle{empty} \vskip 1cm  \noindent
{\Large\bf
Strangeness Correlation: \\
A Clue to Hadron Modification in Dense Matter?
}\\[3mm] 
\def\rightmark{Strangeness Correlation}\def\leftmark{M. Asakawa and T. Cs\"org\H o }
\hspace*{6.327mm}\begin{minipage}[t]{12.0213cm}{\large\lineskip .75em
M. Asakawa$^{1,2a}$ and T. Cs\"{o}rg\H{o}$^{1,3b}$
}\\[2.812mm] 
\hspace*{-8pt}$^1$ Department of Physics, Columbia University\\
New York, NY 10027, U.S.A. \\[0.2ex]
\hspace*{-8pt}$^2$ Department of Physics, School of Science,
Nagoya University\\
Nagoya, 464-01, Japan \\[0.2ex]
\hspace*{-8pt}$^3$ KFKI Research Institute for Particle and Nuclear
Physics\\
H-1525 Budapest, POB 49, Hungary
\\[4.218mm]{\it
Received 11 July 1996
}\\[5.624mm]\noindent
{\bf Abstract.} We discuss the possibility to observe hadron modification
in hot matter via the correlation of identical particles.
We find that the hadron modification causes a new type of correlation,
back-to-back correlation.
\end{minipage}

\section{Introduction}
The Hanbury-Brown Twiss (HBT) effect has been widely measured in heavy ion
collisions. It has been expected that the effect will give some clue to
the size of the system at freeze-out. Another interesting topic in heavy
ion physics is the possibility of hadron modification in medium. So far
these two have been considered as two different aspects of heavy ion physics.
The HBT effect is concerned with freeze-out where interaction disappears,
whereas the hadron modification is caused by interaction. 
However, this picture is purely classical. Since the HBT effect is of quantum
nature, we need quantum mechanical consideration.

In relativistic heavy ion collisions, freeze-out looks rather prompt.
The distribution of final state hadrons is, in most of the cases, almost
exponential \cite{QM95}
and this suggests that the system is almost thermalized up to
some time and then breaks up suddenly~\cite{cc}. 
Motivated by this, we have modeled
freeze-out as follows. The system remains thermalized until freeze-out.
Hadrons are modified due to interaction and their masses are shifted.
As a result, it is dressed pseudo-particles that are thermalized.
Then, freeze-out is assumed to take place suddenly. 
In this paper, we investigate how this picture of freeze-out is
described quantum mechanically and how this scenario is ascertained
through the HBT correlation. In Section 2, we present
the theoretical
framework of our work. In Section 3, we show numerical results and give
some discussion. In Section 4, we conclude and give a perspective. 

\section{Calculation of Correlation Functions}
In this paper, we use the following scalar theory and we ignore the isospin
structure, because it is not essential in our argument. Throughout this
paper, we adopt the mean field approximation for simplicity and for clarity.
In addition, we do not take into account the finiteness of the system.
Calculations for finite systems will be presented elsewhere \cite{ac}.

The theory in the vacuum is given by the following free Lagrangian
${\cal L}_0$:
\begin{equation}
{\cal L}_0 = \frac{1}{2}\partial_{\mu}\phi(x)\partial^{\mu}\phi(x)
-\frac{1}{2}m_0^2 \phi^2(x),
\end{equation}
where $m_0$ is the vacuum mass of the scalar field $\phi(x)$.
The Hamiltonian of this system is, of course, given by
\begin{equation}
\frac{1}{2}\int(\pi^2(x) + \sum_{i}(\partial_i\phi(x))^2 + m_0^2
\phi^2(x))\, d^3x.
\end{equation}
The canonical quantization of this field is normally carried out 
for the canonical variable $\phi({\bf k})$ in
momentum space and its canonical conjugate $\pi({\bf k})$ defined by
\begin{eqnarray}
\phi({\bf k}) & = & \frac{1}{(2\pi)^{3/2}}\int\!\phi({\bf x})
\exp(-i\,{\bf k\!\cdot\! x})\, d^3x, \nonumber \\
\pi({\bf k}) & = & \frac{1}{(2\pi)^{3/2}}\int\!\pi({\bf x})
\exp(-i\,{\bf k\!\cdot\! x})\, d^3x,
\end{eqnarray}
by introducing an annihilation operator $a({\bf k})$ and
a creation operator $a^\dagger ({\bf k})$ as follows:
\begin{eqnarray}
\phi({\bf k}) & = & \frac{1}{\sqrt{2\omega_{\bf k}}}
(a({\bf k}) + a ^\dagger (-{\bf k})), \nonumber \\
\pi({\bf k}) & = & -i \, \sqrt{\frac{\omega_{\bf k}}{2}} \label{adef}
(a({\bf k}) - a ^\dagger (-{\bf k})),
\end{eqnarray}
where $\omega_{\bf k} = \sqrt{m_0^2 + {\bf k}^2}$ and
we have suppressed the time dependence of the operators.
After normal ordering, we get the well-known diagonalized Hamiltonian,
\begin{equation}\label{hfree}
H_0=\int\! \omega_{\bf k}a^\dagger({\bf k})a({\bf k}) d^3 k.
\end{equation}
When the temperature and/or chemical density is non-vanishing and
the mass of the $\phi$ field is shifted, within the mean field
approximation, this is expressed by the following Lagrangian in
medium ${\cal L}_M$:
\begin{equation}\label{lmed}
{\cal L}_M = \frac{1}{2}\partial_{\mu}\phi(x)\partial^{\mu}\phi(x)
-\frac{1}{2}(m_0^2+m_1^2) \phi^2(x).
\end{equation}
The mass shift $\delta M$ is given by
\begin{equation}
\delta M = \sqrt{m_0^2 + m_1^2} - m_0 .
\end{equation}
Let us define the quanta that diagonalize the Lagrangian (\ref{lmed})
as $b({\bf k})$. The point here is that the $b$-quanta are, in general,
different from the $a$-quanta which are the fundamental excitations
in the vacuum. In other words, the $a$-operators do not diagonalize
the Hamiltonian in medium $H_M$. By inserting Eq. (\ref{adef}) into
$H_M$ and carrying out normal ordering, we obtain
\begin{eqnarray}
H_M & = & H_0 + H_1, \nonumber \\
H_1 & = & \frac{m_1^2}{4}\int \!\frac{1}{\omega_{\bf k}}
[ a({\bf k})a(-{\bf k})
+ 2a^\dagger(-{\bf k})a(-{\bf k})+a^\dagger({\bf k})a^\dagger(-{\bf k})]
\, d^3 k. \label{hmed}
\end{eqnarray}
Therefore, in medium, mode ${\bf k}$ and mode ${\bf -k}$ of the $a$-quanta
are mixed. This Hamiltonian (\ref{hmed}) can be exactly diagonalized with
the following Bogoliubov transformation:
\begin{eqnarray}
a^\dagger({\bf k})\!\!\!\!&=&\!\!\!\!
\cosh r_{\bf k}\, b^\dagger({\bf k})+\sinh r_{\bf k}\, b(-{\bf k}), \,\,
a^\dagger(-{\bf k}) = \cosh r_{\bf -k}\,
b^\dagger(-{\bf k}) +\sinh r_{\bf -k}\, b({\bf k}),
\nonumber \\ 
a({\bf k})\!\!\!\!& = & \!\!\!\!\sinh r_{\bf k}\,
b^\dagger(-{\bf k}) +\cosh r_{\bf k}\, b({\bf k}), \,\,
a(-{\bf k}) = \sinh r_{\bf -k}\,
b^\dagger({\bf k}) +\cosh r_{\bf -k}\, b(-{\bf k}), \nonumber \\
&&
\label{bog}
\end{eqnarray}
where $r_{\bf k}$ is a real number which satisfies
the following equation,
\begin{equation}
\tanh 2r_{\bf k} = \frac{-m_1^2}{2\omega_{\bf k}^2 + m_1^2}.
\end{equation}
This gives the exact relationship between the quanta in the vacuum
($a$-quanta) and the quanta in medium ($b$-quanta) in this theory.
With the $b$-operators, the in-medium Hamiltonian (\ref{hmed}) is
diagonalized as
\begin{equation}
H_0=\int\! \Omega_{\bf k}b^\dagger({\bf k})b({\bf k}) d^3 k,
\end{equation}
where $\Omega_{\bf k}$ is given by
\begin{equation}
\Omega_{\bf k} = \sqrt{m_0^2 + m_1^2 + {\bf k}^2}.
\end{equation}
This is almost trivial, since the mass of the $\phi$ field is
shifted to $\sqrt{m_0^2 + m_1^2}$ in medium. However, the important point
here is that it is the $b$-quanta that are thermalized in medium, since the
$b$-operators diagonalize the in-medium Hamiltonian. On the other hand,
the $b$-quanta are not observed experimentally. It is the $a$-quanta
that are observed. Therefore, in calculating final state observables,
we have to evaluate the expectation value of operators defined
in terms of $a$ and $a^\dagger$ operators,
${\cal O}(a, a^\dagger)$
with the density matrix defined in the $b$-basis, $\rho_b$, i.e.,
\begin{equation}
\langle {\cal O}(a, a^\dagger)\rangle
= {\rm Tr}\,\rho_b {\cal O}(a, a^\dagger).
\end{equation}
The calculation of the one and two-particle distribution functions
which are needed to obtain the two-particle correlation is straightforward,
but the Glauber -- Sudar\-shan representation of the thermal density matrix
\cite{gs},
\begin{equation}
\rho_b = \prod_{\bf k} \int\! \frac{d^2\beta({\bf k})}{\pi}
P(\beta({\bf k}))
|\beta({\bf k})\rangle\langle\beta({\bf k})|,
\end{equation}
is useful \cite{aw}.
 Here $|\beta({\bf k})\rangle$ is a coherent state satisfying
$b({\bf k}) |\beta({\bf k})\rangle = \beta({\bf k}) |\beta({\bf k})\rangle$
and $P(\beta({\bf k}))$ is defined by
\begin{equation}
P(\beta({\bf k})) = \frac{1}{f_B(\bf k)}
\exp\left (-\frac{|\beta({\bf k})|^2}{f_B(\bf k)}\right ),
\end{equation}
where $f_B(\bf k)$ is the Bose--Einstein distribution function with
mass $\sqrt{m_0^2 + m_1^2}$.
With this formula, we obtain the one-particle distribution in the final
state,
\begin{equation}
\langle a^\dagger({\bf k})a({\bf k}) \rangle
= \cosh 2r_{\bf k} \, f_{B}({\bf k}) + \sinh^2 \! r_{\bf k} .
\end{equation}
Since we have ignored the finite size effect, the two-particle distribution
function
$\langle a^\dagger({\bf k})a^\dagger({\bf k}')
a({\bf k})a({\bf k}') \rangle $ takes a trivial value
$\langle a^\dagger({\bf k})a({\bf k}) \rangle 
\langle a^\dagger({\bf k}')a({\bf k}') \rangle$
unless ${\bf k}=\pm{\bf k}'$.
For ${\bf k}=\pm{\bf k}'$ cases, we obtain
\begin{eqnarray}
\langle a^\dagger({\bf k})a^\dagger({\bf k})
a({\bf k})a({\bf k}) \rangle \!\!\!\! & = & \!\!\!\!
2 \langle a^\dagger({\bf k})a({\bf k}) \rangle^2 , \nonumber \\
\langle a^\dagger({\bf k})a^\dagger({\bf -k})
a({\bf k})a({\bf -k}) \rangle \!\!\!\! & = & \!\!\!\!
\langle a^\dagger({\bf k})a({\bf k}) \rangle^2 +
f_B ({\bf k}) (f_B ({\bf k})\! + \! 1) \sinh^2 \! 2r_{\bf k}
+ 2 \sinh^4 \! r_{\bf k},  \nonumber \\
\end{eqnarray}
where we have assumed the uniformity of the system, i.e.,
$f_B({\bf k})=f_B(-{\bf k})$ and $r_{\bf k}=r_{\bf -k}$.
From this, we get the two-particle correlation functions
$C_2({\bf k}, \pm{\bf k})$ as follows:
\begin{eqnarray}
C_2({\bf k}, {\bf k}) & = & 2, \nonumber \\
C_2({\bf k}, {\bf -k}) & = & 1 +
\frac{f_B ({\bf k}) (f_B ({\bf k})\! + \! 1) \sinh^2 \! 2r_{\bf k}
+ 2 \sinh^4 \! r_{\bf k}}
{(\cosh 2r_{\bf k} \, f_{BE}({\bf k}) + \sinh^2 \! r_{\bf k})^2} 
\nonumber \\
& \neq & 1 \ .
\end{eqnarray}
In all other cases, the two-particle correlation function is 1.
This result is quite interesting. It implies two important issues,
(i) the intercept of the correlation function
$C_2({\bf k}, {\bf k})$ remains the canonical value even if quanta
in medium are different from those in the vacuum (ii) back-to-back
correlation ($C_2({\bf k}, {\bf -k}) \neq 1$)
is generated by hadron modification. This is caused by the mixing of
${\bf k}$ and ${\bf -k}$ modes due to the mean field effect. In our case,
the mean field carries no momentum. Due to momentum conservation, no
other mixing is possible. Hence, only ${\bf k}$ and ${\bf -k}$ modes get
additional correlation.

\section{Results and Discussion}
The medium effects on the momentum distribution of pions and kaons
are evaluated numerically for static fireballs in Figs. 1 and 2.
These figures indicate that the medium modification
of hadronic masses results in a very small change in the shape  
of the momentum distributions. Fig. 3 indicates the momentum
dependence of $C_2({\bf k}, {\bf -k})$ for pions, assuming a 
freeze-out temperature of $T_f = 120$ MeV and less than 1 \% values
for the mass-shift of pions in medium. Figure 3 indicates
that the deviation of $C_2({\bf k}, {\bf -k})$ from unity is very small
for pions and that the effect is smaller than the current levels of error bars
for the two-particle correlation function. However, the
medium modification of kaons may be observable in the two-particle
correlation function according to Figure 4. The enhancement at large
values of particle momenta in the CMS of the fireball is a strongly
increasing function of the medium modification of the kaonic masses.
The effect is decreasing with increasing values of the freeze-out
temperature $T_f$.

 Our result is essentially different from that of Ref.~\cite{aw} where
over-bunching of identical pions was predicted at small relative momenta,
i.e. $C_2({\bf k}, {\bf k}) > 2$, while in our case the intercept parameter
 takes up the canonical thermal value of 2.
 We find, in contrast to Ref.~\cite{aw}, that 
mode-mode mixing generates positive correlation of kaons at
large relative momenta, if the mass of the kaons is modified in the medium.

\begin{center}
\vspace*{8.0cm}
\includegraphics{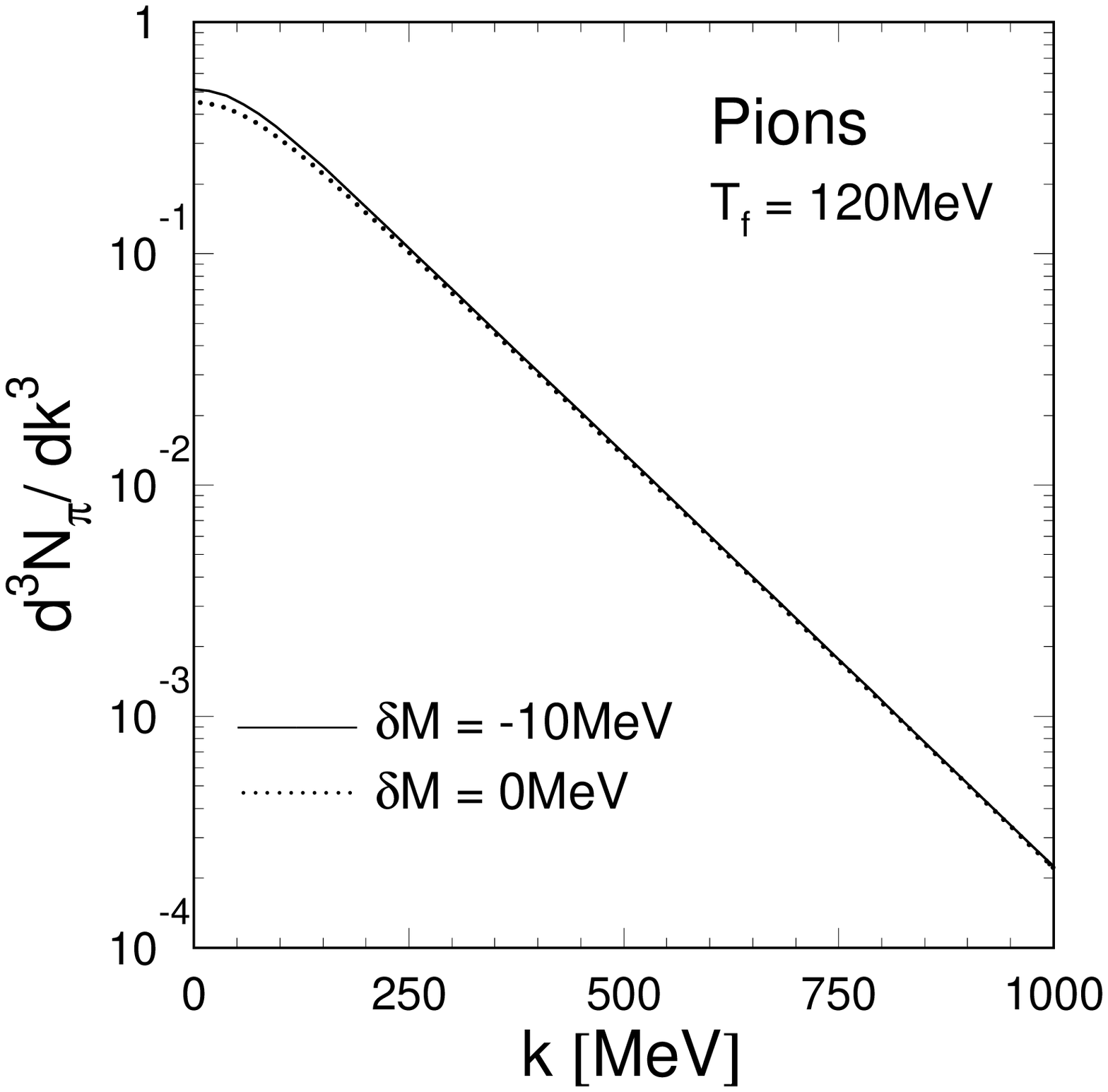}
%\vskip -70pt
%\end{center}
%\begin{center}
\begin{minipage}[t]{11.054cm}
{\small {\bf Fig.~1.}
Solid line indicates the spectrum of pions from a fireball,
when a small in-medium mass-shift of --10 MeV is invoked,
dotted line stands for the no mass-shift case.}
\end{minipage}
\end{center}
\vskip 4truemm
 
\begin{center} 
\vspace*{7.cm}
\includegraphics{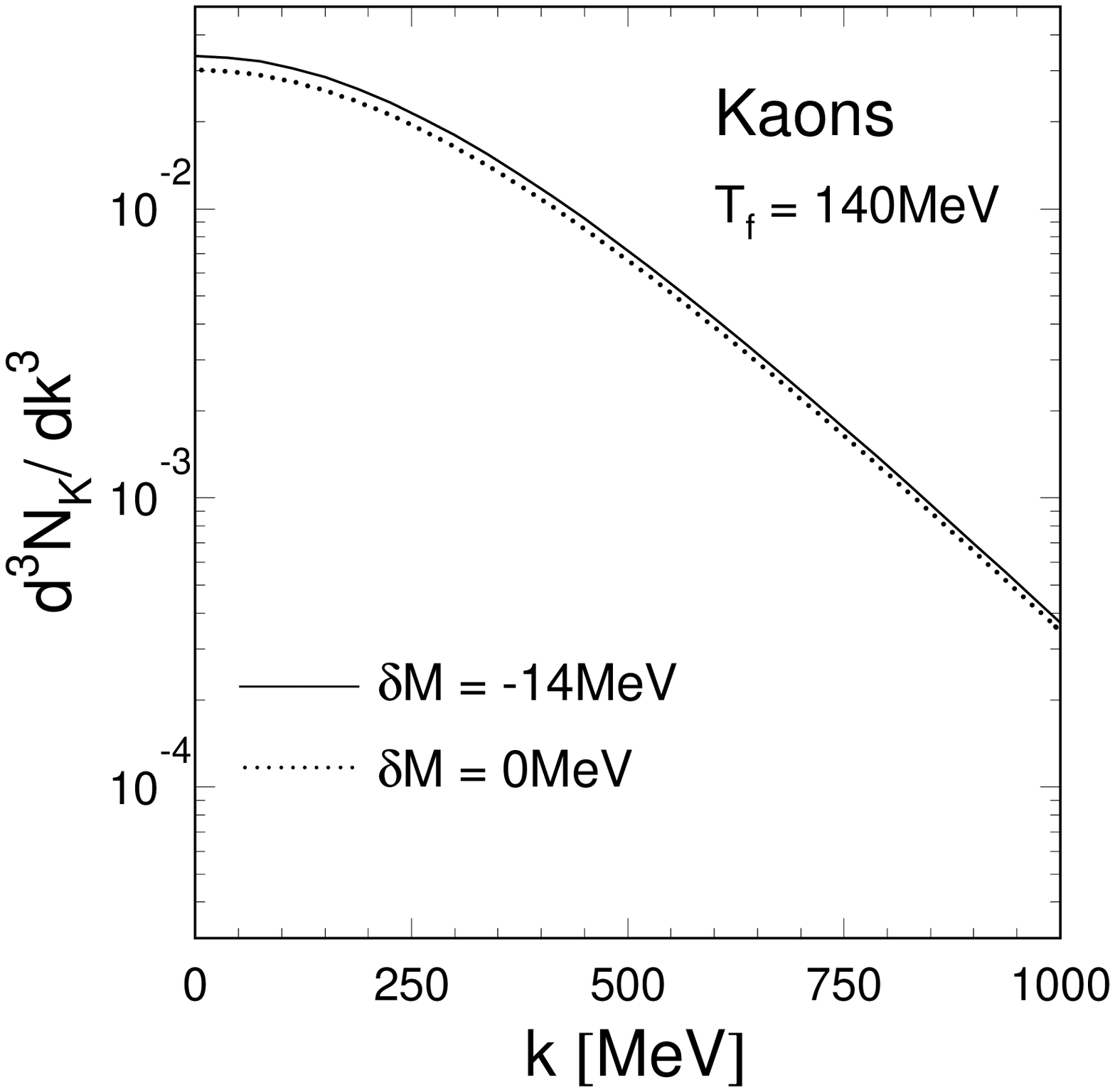}
%\vskip -70pt
%\end{center}
%\begin{center}
\begin{minipage}[t]{11.0540cm}
{\small {\bf Fig.~2.}
Solid line indicates the spectrum of kaons from a fireball,
when a small in-medium mass-shift of --14 MeV is invoked,
dotted line stands for the no mass-shift case.}
\end{minipage}
\end{center}
\vskip 4truemm

\begin{center}
\vspace*{7.5cm}
\includegraphics{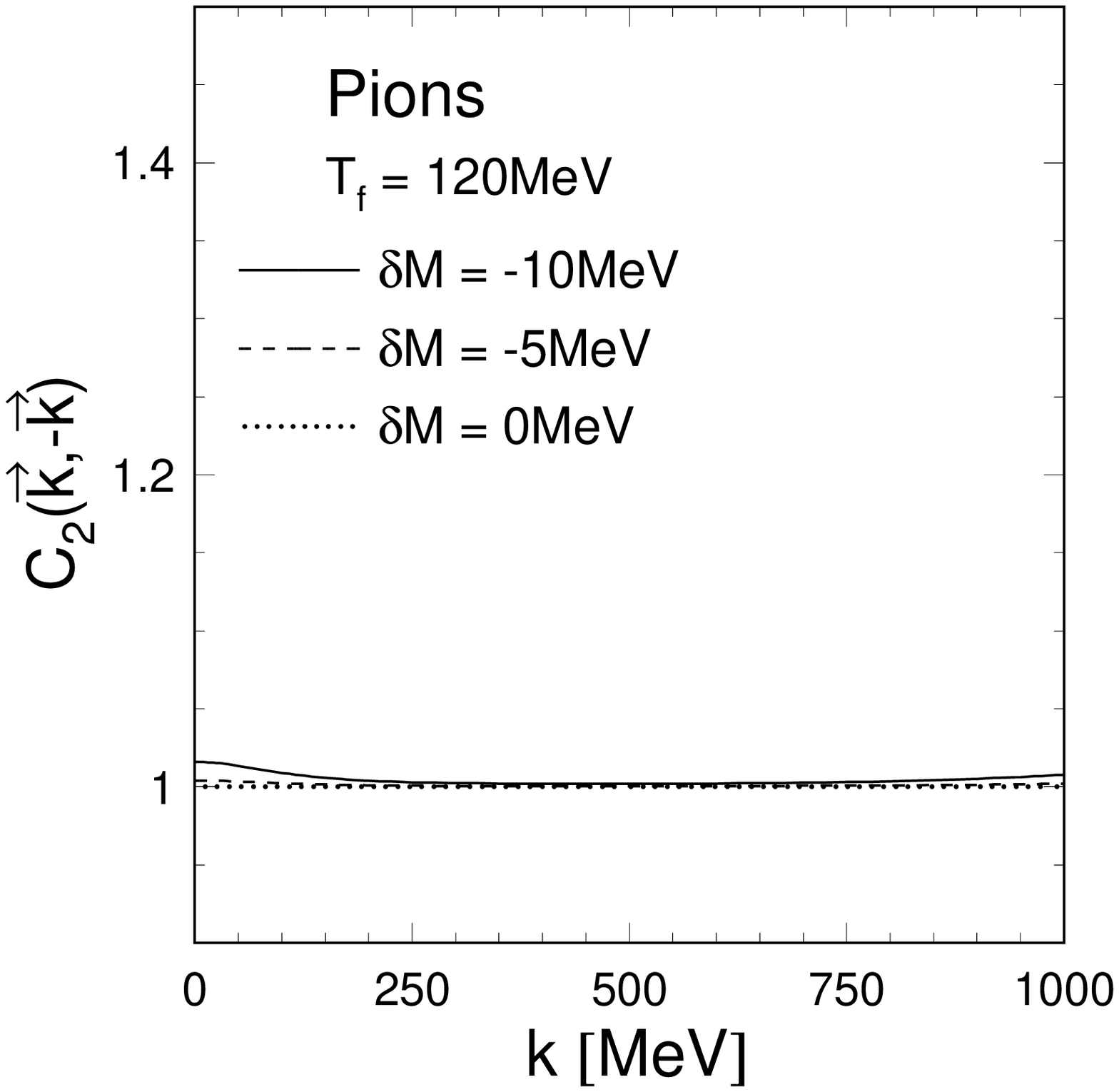}
%\vskip -70pt
%\end{center}
%\begin{center}
\begin{minipage}[t]{11.0540cm}
{\small {\bf Fig.~3.}
Solid line indicates the correlation function $C_2({\bf k}, {\bf -k})$
of pions from a fireball,
when a small in-medium mass-shift of --10 MeV is invoked,
dashed line stands  for the $\delta M = -5$ MeV,
dotted line stands for the no mass-shift case.}
\end{minipage}
\end{center}
\vskip 4truemm

\begin{center}
\vspace*{7.5cm}
\includegraphics{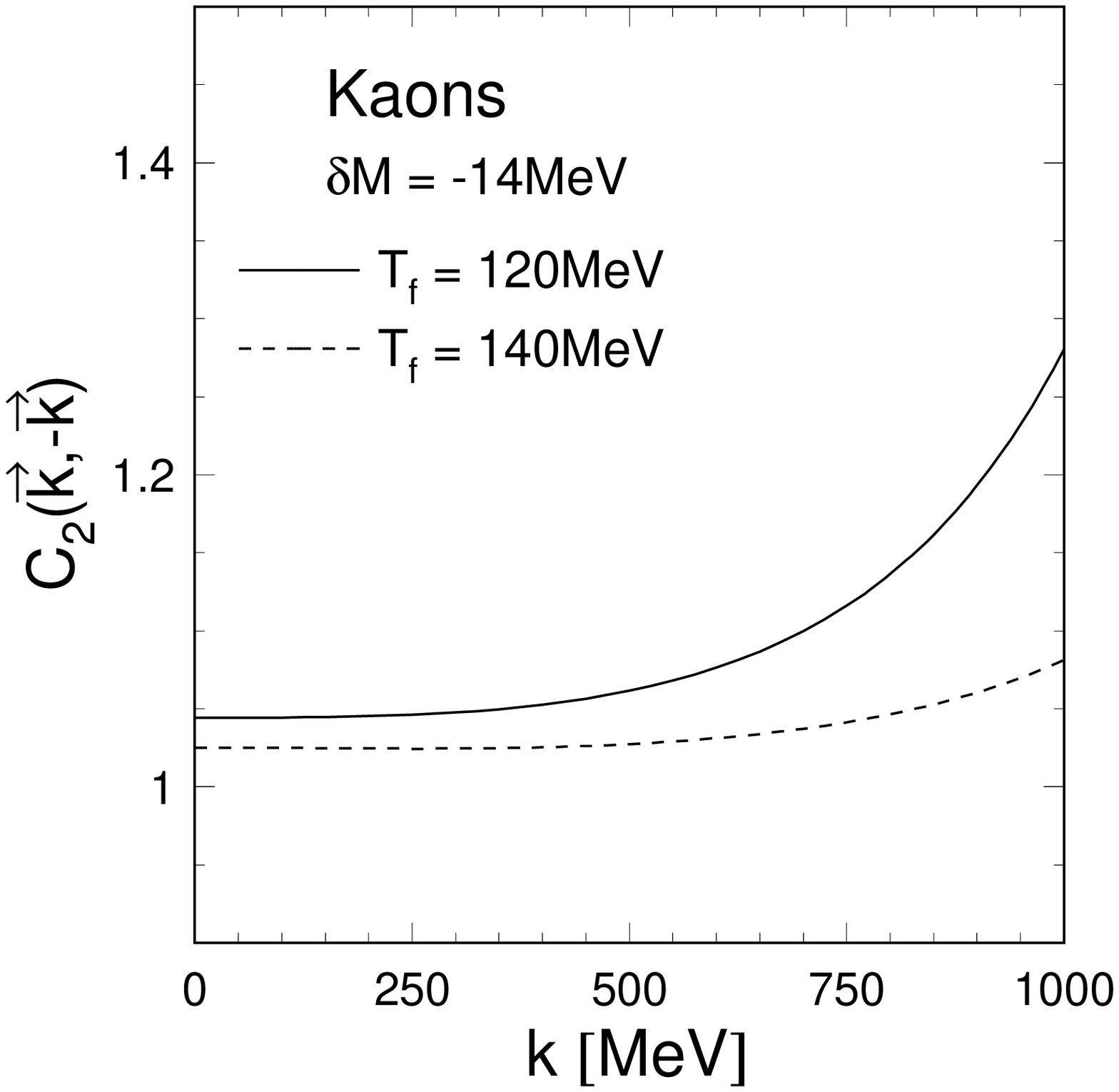}
%\vskip -70pt
\begin{minipage}[t]{11.0540cm}
{\small {\bf Fig.~4.}
Illustration of the dependence of $C_2({\bf k}, {\bf -k})$
on the value of the \mbox{freeze-out} temperature $T_f$, for kaons.}
\end{minipage}
\end{center}
\vskip 4truemm

\begin{center}
\vspace*{8.0cm}
\includegraphics{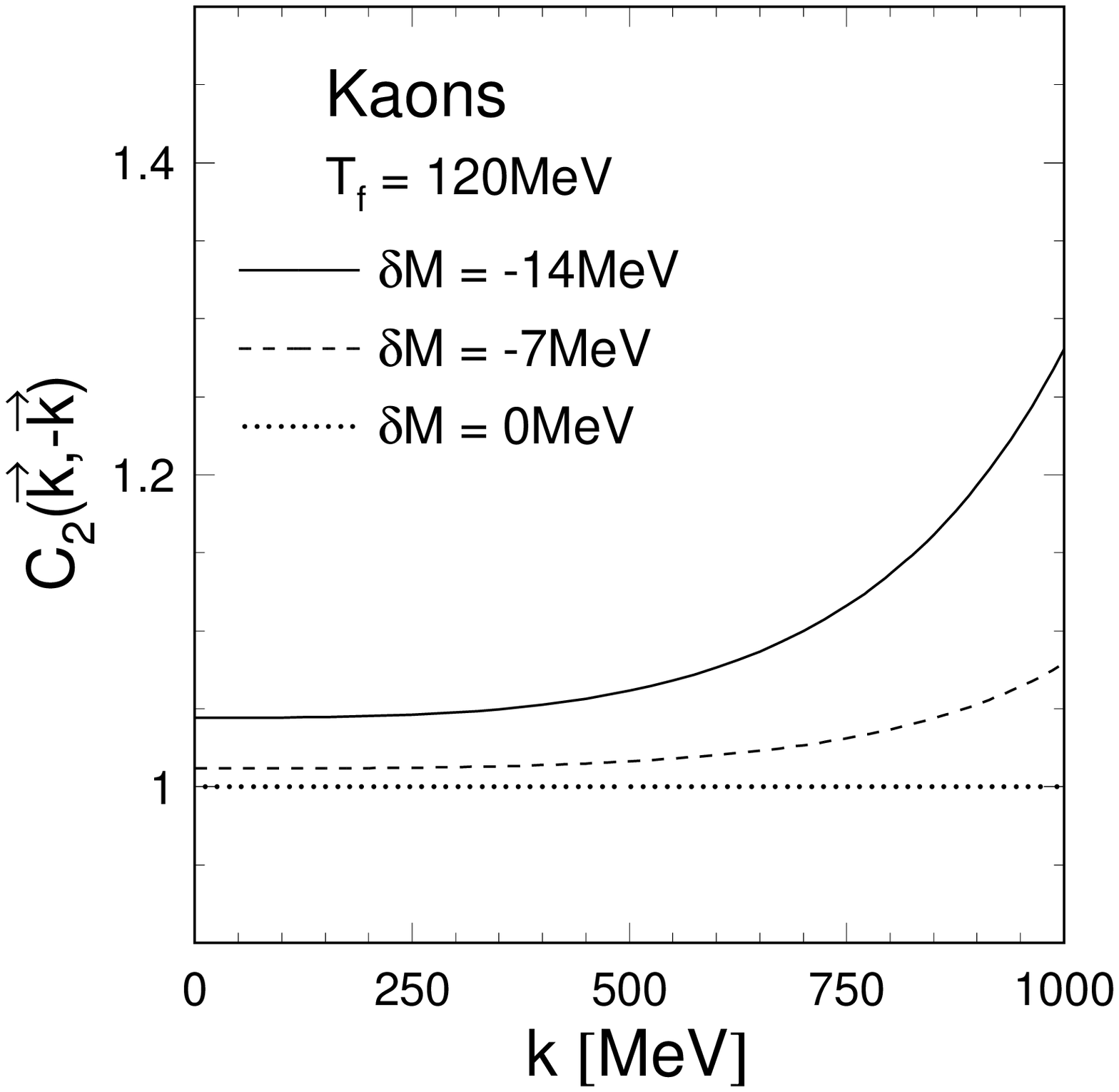}
%\vskip -70pt
\begin{minipage}[t]{11.054cm}
{\small {\bf Fig.~5.}
Same as Fig. 3, but for kaons, and a little larger mass-shifts.}
\end{minipage}
\end{center}
\vskip 10pt
\vfill\eject

\section{Summary}

 A new method has been found to test the medium modification of
kaons, utilizing their quantum correlations at large
momentum difference. The effect follows from basic principles
of statistical physics and canonical quantization.

 Back-to-back correlations are not contaminated by resonance decays, in
contrast to the Bose--Einstein correlation functions at small
relative momenta.
For a locally thermalized expanding source, however,  
back-to-back correlations will appear in the rest frame 
of each fluid element. Thus, for such  
systems a more realistic estimate is necessary to learn more
about the magnitude of the correlations of kaons at large relative momenta.

In the present study we have 
neglected finite size effects, which will make the 
correlation function vary smoothly around both ${\bf k_1} = {\bf k_2}$
and ${\bf k_1} =  - {\bf  k_2}$.

\vskip 15pt

%\noindent \large {\bf 
%\normalsize
%\vskip 10pt
%
\section*{Acknowledgment}
\noindent
We would like to thank M. Gyulassy and J. Knoll for enlightening
discussions.  \mbox{T. Cs\"org\H o} is grateful to  Gy\"orgyi and M. Gyulassy for
kind hospitality at Columbia University.
This research has been supported by the US - Hungarian Joint Fund
under contract MAKA 378/1993, by the Hungarian NSF under contract
OTKA - F 4019 and by the Advanced Research Award of the  Fulbright Foundation,
grant document 20926/1996. 

\section*{Notes}
\begin{itemize}
\item[\null]
a. E-mail: yuki@nuc-th.phys.nagoya-u.ac.jp
\item[\null]
b. E-mail: csorgo@sunserv.kfki.hu
\end{itemize}

\vfill\eject 

\vskip 15pt

\begin{thebibliography}{99}\parindent=8truemm
\itemsep -1mm

\bibitem{QM95}{\it Quark Matter '95}, {\it Proceedings of the 11th International
Conference on Ultra-Relativistic Nucleus-Nucleus Collisions},
{\it Nucl. Phys.} {\bf A590} (1995) Nos. 1, 2.

\bibitem{cc} T. Cs\"org\H o and L. P. Csernai, {\it Phys. Lett.} {\bf B333}
(1994) 494.

\bibitem{ac} M. Asakawa and T. Cs\"{o}rg\H{o}, in preparation.

\bibitem{gs} R. J. Glauber, {\it Phys. Rev.} {\bf 131} (1963) 2766;
E.C.G. Sudarshan, {\it Phys. Rev. Lett.} {\bf 10} (1963) 277.

\bibitem{aw} I. Andreev and R. M. Weiner, {\it  Phys. Lett.} {\bf B373} (1996) 
159.
 
\end{thebibliography}
\end{document}